\pdfoutput=1
\documentclass[aps,pre, twocolumn, groupedaddress]{revtex4-1}
\usepackage{lipsum}
\usepackage{float}
\usepackage{mathtools}
\usepackage{graphicx}
\usepackage{subfigure}
\usepackage{caption}
\usepackage{dcolumn}
\usepackage{amsmath}    
\usepackage{amssymb}
\usepackage{amsmath}
\usepackage{bm}
\usepackage{hyperref}
\usepackage{latexsym}
\usepackage{verbatim}
\usepackage{enumitem}
\usepackage[normalem]{ulem}
\usepackage{color}
\usepackage{parskip}
\usepackage{url}
\def\beq{\begin{equation}}
\def\eeq{\end{equation}}

\def\beq{\begin{equation}}                          
\def\eeq{\end{equation}}                          
\def\bea{\begin{eqnarray}}                          
\def\eea{\end{eqnarray}}

\DeclareRobustCommand{\uvec}[1]{{%
  \ifcsname uvec#1\endcsname
     \csname uvec#1\endcsname
   \else
    \bm{\hat{\mathbf{#1}}}%
   \fi
}}
\preprint{}
\begin{document}
\preprint{}

\title {Shape Evolution and Dynamics of Deformable Ring}

\author{Arun Kumar$^{1}$}
\email{arunkumar.rs.phy23@itbhu.ac.in}
\author{Partha Sarathi Mondal$^{1}$}
\email{parthasarathimondal.rs.phy21@itbhu.ac.in}
\author{Pritha Dolai$^{2}$}
\email{pritha@nitk.edu.in}
\author{Shradha Mishra$^{1}$}
\email{smishra.phy@iitbhu.ac.in}
\affiliation{
\textsuperscript{1}Department of Physics, Indian Institute of Technology (BHU), Varanasi, India 221005\\
\textsuperscript{2}Department of Physics, National Institute of Technology Karnataka, Surathkal, India 575025
}

\date{\today}

\begin{abstract}
    {We numerically investigate the dynamics of a deformable closed ring filled with active particles. The ring is modeled as a flexible boundary made up of passive beads interacting with a harmonic spring force. The interior of the ring is filled with active Brownian particles (ABPs),  and their activity is controlled through the rotational diffusion coefficient. We explore how, by systematically varying the activity of ABPs,  packing fraction, and size of the ring,  we can control the shape deformation and dynamics of the ring. At low packing fractions, low rotational diffusion coefficients, and smaller ring sizes, the ring exhibits highly irregular and strongly deformed shapes due to the uneven spatial arrangement of active particles along the boundary. Increasing the packing fraction, rotational diffusion coefficient, or ring size promotes a more even distribution of active particles within the ring, thereby suppressing shape deformations and fluctuations, driving the ring toward a more circular shape. We further analyze the mean-squared displacement (MSD) of the ring's center of mass and observe a crossover from ballistic to diffusive dynamics, which can be tuned by varying the system parameters. Our results demonstrate that, despite its internal complexity and deformability, the ring exhibits emergent behavior analogous to that of a single effective active particle. This study provides insight into the collective effects of confined active matter and the resulting macroscopic dynamics of deformable systems.}
    \end{abstract}
 \maketitle
\section{Introduction \label{Introduction}}

Active matter comprises systems of self-driven constituents that continuously consume energy to generate motion and mechanical forces \cite{bechinger2016active,semwal2024dynamics,ramaswamy2010mechanics}. These systems span a wide range of biological and synthetic realizations, including bacterial colonies, cytoskeletal networks, epithelial tissues, animal collectives, and artificial microswimmers \cite{htet2025analytical}. Unlike passive systems at equilibrium, active matter operates far from thermodynamic equilibrium, enabling sustained energy injection at the microscopic scale. As a result, these systems exhibit a wide variety of emergent behaviors, many of which have no direct equilibrium analog \cite{wu2000particle,mishra2023active}.

Despite their non-equilibrium nature\cite{o2022time}, active systems often display structural organization reminiscent of conventional phases such as gases, liquids, and solids. However, their dynamics are fundamentally distinct, as activity can drive spontaneous flows, long-range correlations, and persistent collective motion\cite{marchetti2013hydrodynamics,giomi2011polar,ramaswamy2006mechanics}. One of the defining features of active matter is the ability of individual agents to convert stored or ambient energy into directed motion, which, when coupled across many particles, leads to large-scale organization and pattern formation \cite{fily2012athermal}.

A key concept underlying these phenomena is the emergence of collective mechanical forces generated by active constituents\cite{ramaswamy2010mechanics,kushwaha2026emergent}. These forces originate from microscopic force production by individual active units and become organized at macroscopic scales through interactions and alignment mechanisms. In biological systems, such mechanically generated forces play a central role in processes including intracellular transport, cytoskeletal remodeling, morphogenesis, and cell migration \cite{blanchoin2014actin,balasubramaniam2022active}. Importantly, these processes typically occur within confined and deformable environments, where mechanical feedback between internal activity and boundary properties plays a crucial role \cite{htet2025analytical,peterson2021vesicle}.

Confinement has been shown to influence the behavior of active systems significantly\cite{duzgun2018active}. In contrast to equilibrium systems, where boundary effects are usually localized near interfaces, confinement in active matter can affect the entire system by redistributing stresses and particle organization\cite{duzgun2018active,thutupalli2018flow,vsindelka2025confined,paoluzzi2015self}. This can give rise to a range of phenomena, including rectification, density inhomogeneities, spontaneous flow generation, and collective oscillations\cite{dolai2018phase}. The presence of confinement thus introduces an additional level of control over the emergent dynamics of active systems.

When the confining boundary is deformable, the system exhibits an even richer set of behaviors due to the presence of feedback between active forces and boundary mechanics\cite{peterson2021vesicle,tian2015boundary,uplap2023design}. Active stresses can deform the boundary, altering its shape and curvature, while the resulting geometry influences particle accumulation, orientation, and transport. This bidirectional coupling can lead to symmetry breaking, polarity formation, and directed motion of the entire confined system, as well as complex shape transformations \cite{paoluzzi2016shape,hiraiwa2010dynamics,iyer2023dynamic,baruffi2019overdamped,peng2022activity,quillen2020boids,uplap2023design,lee2023complex,chen2017rotational}. Such effects are particularly relevant in biological contexts, where rings are inherently flexible and responsive.

Most previous studies have focused either on active particles confined by rigid boundaries or on systems described using continuum theories that impose specific forms of internal order\cite{fily2012athermal}. While these approaches have provided valuable insights, they often rely on assumptions that limit the emergence of structure and dynamics from microscopic interactions\cite{spellings2015shape,diaz2024active}. In contrast, the self-consistent emergence of both internal organization and boundary deformation in minimal particle-based models has been comparatively less explored \cite{king2026active,tian2015boundary,peterson2021vesicle,hadjifrangiskou2025nematic,ohta2009deformable,ohta2009deformation}.

In this work, we investigate a minimal model of active Brownian particles (ABPs) confined within a flexible ring\cite{nikola2016active}. By combining particle interactions with ring mechanics, we identify a generic feedback mechanism between particle dynamics and ring deformation. This interplay drives spontaneous symmetry breaking, shape remodeling, and persistent motion of the ring as a whole\cite{paoluzzi2016shape}. Our results demonstrate that complex, life-like behaviors can emerge from simple mechanical principles, providing new insight into biological organization and offering design guidelines for synthetic active materials and artificial cell-like systems \cite{ohta2009deformable,ohta2009deformation,deschamps2009dynamics,knippenberg2024motility}.

The remainder of this paper is organized as follows. In Sec.~II, we describe the details of the model. In Sec.~II~A, we present the ring dynamics, while Sec.~II~B discusses the Active Particle dynamics. In Sec.~III, we present the results on ring shape and its transport properties. Specifically, ring shape is analyzed followed by the asphericity in Sec.~III~A and the active ring motion in Sec.~III~B. In Sec.~IV, we demonstrate that the ring can be effectively described as an effective Brownian particle. Finally, the conclusions are summarized in Sec.~V.

\begin{figure*}[hbtp]
    \centering
    \includegraphics[width=1.0\linewidth]{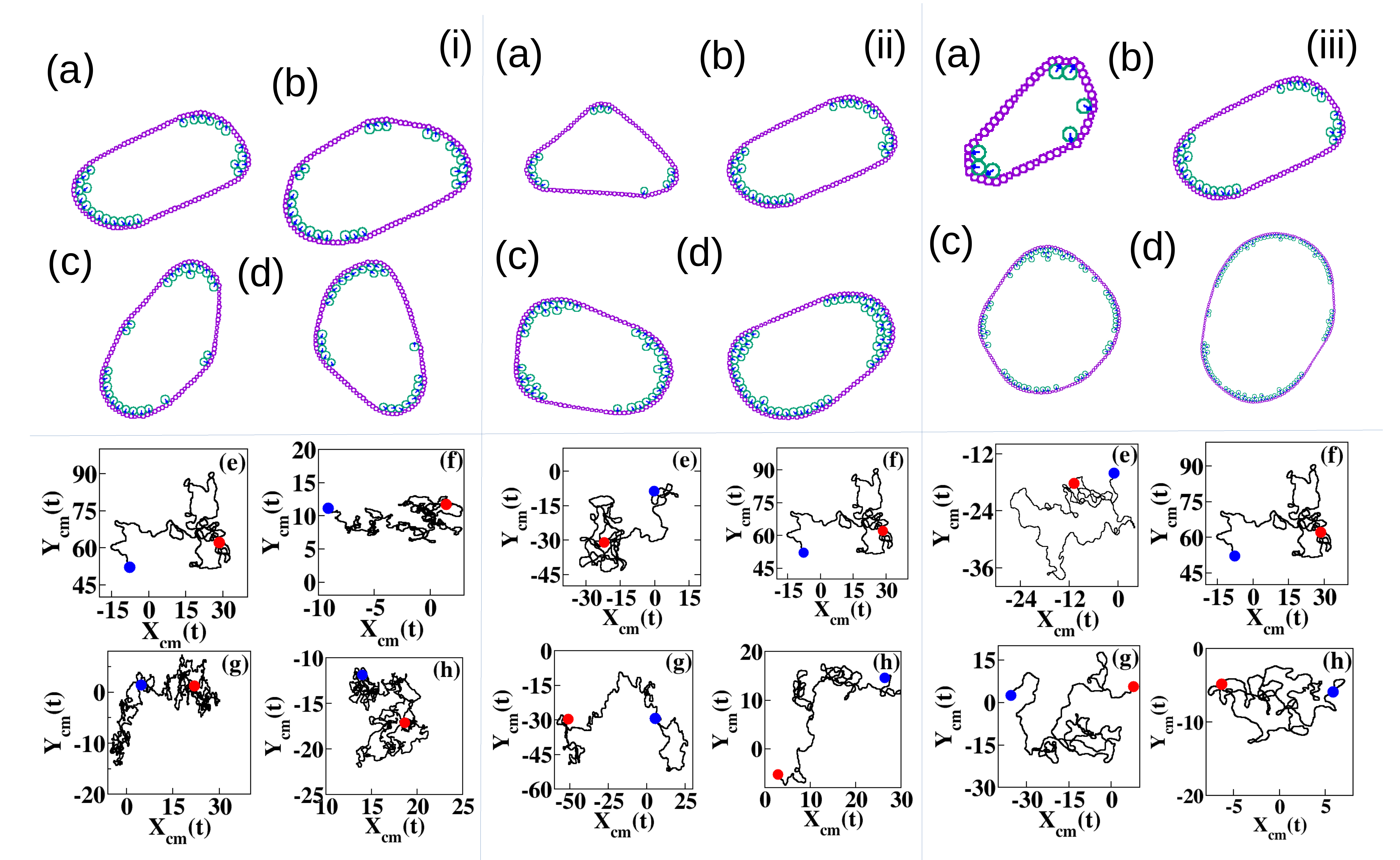}
       \caption {Shape transformations of a deformable ring under different system parameters. Configurations and corresponding center of mass trajectories: \textbf{(i)} fixed $N_{\mathrm{beads}} = 100$, $\phi = 0.10$, and for different $D_r = 0.01$, $0.03$, $0.05$, and $0.07$, shown in (a-d), (e-h), respectively. \textbf{(ii)} fixed $N_{\mathrm{beads}} = 100$, $D_r = 0.10$, and for different $\phi = 0.05$, $0.10$, $0.15$, and $0.20$, shown in (a-d), (e-h), respectively. \textbf{(iii)} fixed $D_r = 0.01$, $\phi = 0.1$, and for different $N_{\mathrm{beads}} = 50$, $100$, $150$, and $200$, shown in (a-d), (e-h), respectively.  The magenta and cyan circles in (a-d) in all three (i)-(iii) columns represent the beads and ABPs, respectively. The red and blue solid circles represent the start and end point of a single trajectory. All trajectories are plotted for equal time.}
    \label{fig:fig1}
\end{figure*}
\section{Model and Method \label{Model and Method}}
We have a single deformable ring made up of interacting beads, and it is filled with active Brownian particles (ABPs). The whole system is kept in an open two-dimensional plane. Hence, the model is divided into two main parts: bead particle dynamics and active particle dynamics. We perform simulations for $N_{\mathrm{beads}}$ bead particles of radius  $r_b$ and   $N_{\mathrm{active}}$ active Brownian particles of radius  $r_a$.

\subsection{Bead Particle Dynamics}

We model the boundary as a closed ring of $N_{\mathrm{beads}}$  made up of colloidal beads,  connected by elastic springs \cite{paoluzzi2016shape}. The time evolution of the position $\mathbf{R}_n$ of the $n$-th bead is governed by the overdamped Langevin's dynamics
\begin{equation}
\frac{d\mathbf{R}_n}{dt} = \mu \, \mathbf{F}_{\mathrm{beads}},
\label{overdamped}
\end{equation}
where $\mu$ is the mobility, and $\mathbf{F}_{\mathrm{beads}}$ is the total force acting on bead $n$, given by
\begin{equation}
\mathbf{F}_{\mathrm{beads}}= \mathbf{F}_{\mathrm{harm}} + \mathbf{F}_{\mathrm{rep}}
\end{equation}
\begin{equation}
    \mathbf{F}_{\mathrm{harm}} = - \boldsymbol{\nabla} V_{\mathrm{harm}}(\{\mathbf{R}\})
\end{equation}

The harmonic potential enforces connectivity between neighboring beads and is given by
\begin{equation}
V_{\mathrm{harm}}(\{\mathbf{R}\}) 
= \frac{k_{\mathrm{harm}}}{2} \sum_{n=1}^{N_b} 
\left( |\mathbf{R}_{n+1} - \mathbf{R}_n| - 2r_b \right)^2,
\end{equation}
where $k_{\mathrm{harm}}$ is the spring constant and $2r_b$ is the bead diameter or the equilibrium bond length.

The repulsive interaction accounts for excluded-volume effects and includes both bead-bead and bead--particle interactions,
\begin{equation}
V_{\mathrm{rep}}(\{\mathbf{R}\}, \{\mathbf{r}\}) 
= \sum_{n<m} V(|\mathbf{R}_n - \mathbf{R}_m|) 
+ \sum_{n,i} V(|\mathbf{R}_n - \mathbf{r}_i|).
\end{equation}

\begin{equation}
\mathbf{F}_{\mathrm{rep}} = - \nabla V_{\mathrm{rep}}(\{\mathbf{R}\}, \{\mathbf{r}\}).
\end{equation}
\begin{equation}
    \mathbf{F}_{\mathrm{rep}} = \mathbf{F}_{\mathrm{rep}}^{\mathrm{bb}} +\mathbf{F}_{\mathrm{rep}}^{\mathrm{ba}}
\end{equation}
Where $\mathbf{F}_{\mathrm{rep}}^{\mathrm{bb}}$ is the repulsive force between the bead particles and $\mathbf{F}_{\mathrm{rep}}^{\mathrm{ba}}$ is the repulsive force between the beads and the active particle. 

\begin{equation}
\mathbf{F}_{\mathrm{rep}}^{\mathrm{bb}} =  
\begin{cases}
k_{\mathrm{rep}}  \left( 2r_b - |\mathbf{R}_n - \mathbf{R}_m| \right) 
\hat{\mathbf{e}}_{nm}, & \text{if } |\mathbf{R}_n - \mathbf{R}_m| < 2r_b, \\
0, & \text{otherwise},
\end{cases}
\end{equation}

where $\hat{\mathbf{e}}_{nm} = (\mathbf{R}_n - \mathbf{R}_m)/|\mathbf{R}_n - \mathbf{R}_m|$ is the unit vector along the line joining the two particles. 

The interaction potential $V_{\rm{rep}}$ is a short-range repulsive potential that acts only when two particles overlap. Consequently, the force experienced by an active particle located at $\mathbf{r}_i$, due to its interaction with the bead particles is obtained by summing the pairwise repulsive forces exerted by all beads. Specifically, the pairwise interaction force between a bead located at $\mathbf{R}_n$ and an active particle at $\mathbf{r}_i$ is given by
\begin{equation}
\mathbf{F}_{\mathrm{rep}}^{\mathrm{ba}} =  
\begin{cases}
k_{\mathrm{rep}}  \left( 2s - |\mathbf{R}_n - \mathbf{r}_i| \right) 
\hat{\mathbf{e}}_{ni}, & \text{if } |\mathbf{R}_n - \mathbf{r}_i| < 2s, \\
0, & \text{otherwise},
\end{cases}
\end{equation}
where $\hat{\mathbf{e}}_{ni} = (\mathbf{R}_n - \mathbf{r}_i)/|\mathbf{R}_n - \mathbf{r}_i|$ is the unit vector along the line joining the two particles.

Thus, interactions are purely repulsive and act only when the interparticle distance is smaller than the effective diameter $2s$, ensuring excluded-volume constraints. $2s = (r_b+r_a)$ is the sum of the radius of the bead and the active particle. 

\subsection{Active Particle Dynamics}

The dynamics of the Active particles are modeled using overdamped Langevin dynamics. Since the active particles are self-propelled, the Langevin equations include both an active velocity and an orientation vector to describe their self-driven motion \cite{bechinger2016active}.  

We have ${N_{\mathrm{active}}}$ active Brownian particles (ABPs) of radius $r_a$, confined by the deformable ring of bead particles. The position of the $i$-th particle, $\mathbf{r}_i$, evolves according to
\begin{equation}
\frac{d\mathbf{r}_i}{dt} = v_0 \mathbf{p}_i(t) + {\mu_a} \,\mathbf{F}_{\mathrm{active}},
\end{equation}
where  $\mathbf{F}_{\mathrm{active}}$ is the total force acting on Active particle and $\mu_a$ is the mobility of active particles. 
\begin{equation}
     \mathbf{F}_{\mathrm{active}} =  \mathbf{F}_{\mathrm{rep}}^{\mathrm{ab}} +\mathbf{F}_{\mathrm{rep}}^{\mathrm{aa}}
\end{equation}
where $\mathbf{F}_{\mathrm{rep}}^{\mathrm{ba}}$ = -$\mathbf{F}_{\mathrm{rep}}^{\mathrm{ab}}$ denotes the repulsive force between beads and active particles, as defined in Eq.~(9), and $\mathbf{F}_{\mathrm{rep}}^{\mathrm{aa}}$ denotes the active-active force defined, 
\begin{equation}
  \mathbf{F}_{\mathrm{rep}}^{\mathrm{aa}} =  
\begin{cases}
k_{\mathrm{rep}} \left( 2r_a - |\mathbf{r}_i - \mathbf{r}_j| \right) 
\hat{\mathbf{e}}_{ij}, & \text{if } |\mathbf{r}_i - \mathbf{r}_j| < 2r_a, \\
0, & \text{otherwise},
\end{cases}
\end{equation}
where $\hat{\mathbf{e}}_{ij} = (\mathbf{r}_i - \mathbf{r}_j)/|\mathbf{r}_i - \mathbf{r}_j|$ is the unit vector along the line joining the two particles.

The orientation of the $i$-th particle is defined as 
\begin{equation}
\frac{d\theta_i}{dt}=\sqrt{2D_r}\,\boldsymbol{\zeta}_{\theta}.
\end{equation}
where the direction vector is 
\[
\mathbf{p}_i(t) = \bigl(\cos \theta_i(t), \, \sin \theta_i(t)\bigr),
\] 
\(D_r\) represents the rotational diffusion coefficient that determines how rapidly the particle changes its direction due to rotational fluctuations. The term \(\zeta_{\theta}(t)\) is an independent Gaussian white-noise satisfying $\langle \zeta_{\theta}(t) \rangle = 0$ and
$\langle \zeta_{\theta}(t)\zeta_{\theta}(t') \rangle = \delta(t-t')$. Here, \(\delta(t-t')\) denotes the Dirac delta function. The parameter \(v_0\) denotes the constant self-propulsion speed of the active particle along its instantaneous orientation direction. Together, these quantities describe the persistent random motion of an ABP, where self-propulsion drives directed motion while rotational diffusion gradually randomizes the particle orientation over time.

\begin{figure*}[hbtp]
    \centering
    \includegraphics[width=1.0\linewidth]{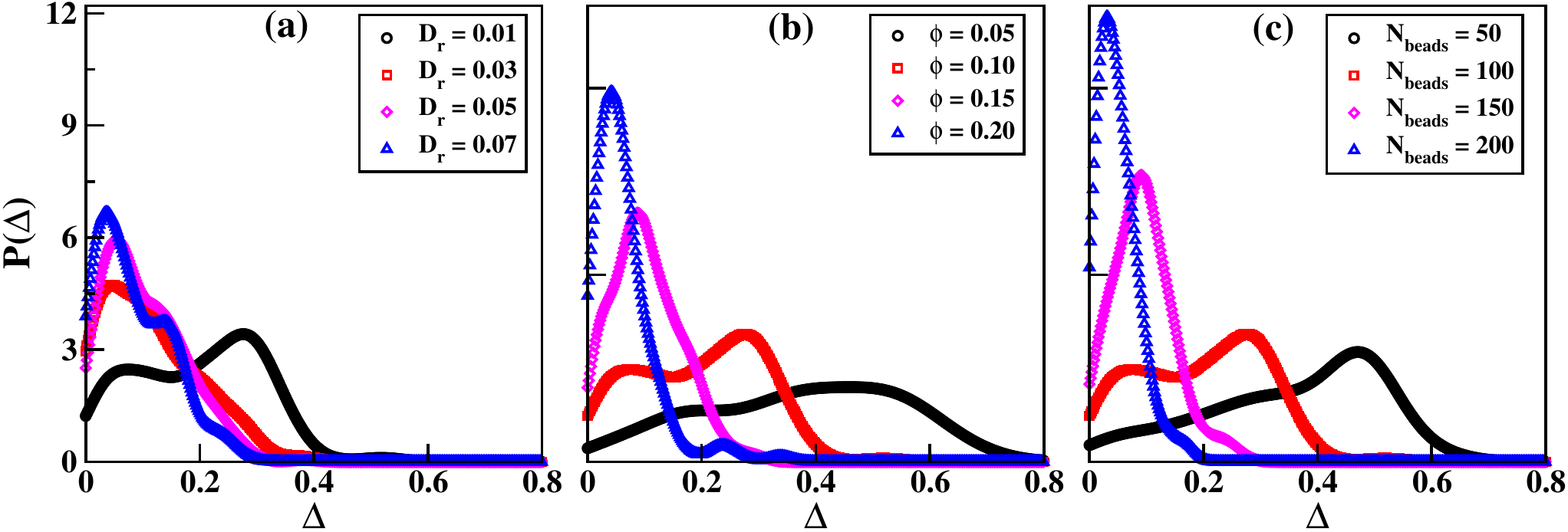}
   \caption{Probability distribution $P(\Delta)$ of the asphericity parameter $\Delta$. \textbf{(a)} At fixed $N_{\mathrm{beads}} = 100$, $\phi = 0.10$ and  for different  $D_r = 0.01$, $0.03$, $0.05$, and $0.07$. \textbf{(b)} At fixed $N_{\mathrm{beads}} = 100$, $D_r = 0.10$, and for different  $\phi = 0.05$, $0.10$, $0.15$, and $0.20$. \textbf{(c)} At  fixed $D_r = 0.01$, $\phi = 0.1$ and for different $N_{\mathrm{beads}} = 50$, $100$, $150$, and $200$. }
 \label{fig:probe}
\end{figure*}

\subsection{Simulation details and dimensionless parameters}
The initial configuration of the ring is characterized by a circle of radius $R_0=\frac{N_{\mathrm{beads}}2 r_b}{2\pi}$. The corresponding circular ring area is given by $A_{\mathrm{beads}}=\pi R_0^2$. The beads are initially positioned on the circumference of the circle according to the parametric equations $x = R_0\cos\psi, y = R_0\sin\psi$, where $\psi$ is the angular coordinate. Active particles are then randomly distributed within the enclosed circular ring. The packing fraction of active particles inside the ring is defined as $\phi  = \frac{N_{\mathrm{active}} \, \pi r_a^2}{A_{\mathrm{beads}}}$.

 We performed numerical simulations of rings composed of $N_{\mathrm{beads}}$ = $50$, $100$, $150$, and $200$ beads enclosing active particles with packing fractions $\phi = (0.05 - 0.20)$. The rotational diffusion coefficient is varied over the range $D_r = (0.01 - 0.07)$. The size of active particles is double the size of the bead particles. We fix the radius of the active particle $r_a = 0.1$ and that of the bead particles $r_b = 0.05$. Further, we set the intrinsic length and time scale in our simulation by $r_a$ and $\tau'= r_a/v_0 = 0.2$, respectively.  The self-propulsion speed of the active particles was fixed at $v_0 = 0.5$. The mobilities of the bead and active particles are set to $\mu=5.0$ and $\mu_a=1.0$, respectively, while the harmonic and repulsive spring constants are $k_{\mathrm{harm}}=500$ and $k_{\mathrm{rep}}=100$. These parameters determine the characteristic elastic relaxation times associated with the harmonic and repulsive interactions, given by $\tau_{\mathrm{harm},e} =\left(\mu k_{\mathrm{harm}}\right)^{-1} =2\times10^{-3}\tau'$, and $\tau_{\mathrm{rep},e} =\left(\mu k_{\mathrm{rep}}\right)^{-1} =5\times10^{-2}\tau'$ respectively. The corresponding persistence length is given by $l=\frac{v_0}{D_r}$, which varies in the range  $l = (500-70)r_a$  for $D_r = (0.01-0.07)$.  Thus, $D_r$ serves as a control parameter governing the persistence of active particle motion. By varying $D_r$, we are tuning the mean time required for one full rotation, which lies in the range $D_r^{-1} = (72.5-500)\tau'$. The initial radius of the ring $R_0$ changes from $\sim (8-32)r_a$, by varying $N_{\mathrm{beads}}$.

The equations of motion were integrated using the Euler–Maruyama algorithm \cite{platen1992numerical} with time step $\mathrm{d}t = 5 \times 10^{-4} \tau'$. All simulations were performed for a total time $T = 4.5 \times 10^{4}\,\tau'$. The first $10^7$ steps, corresponding to an equilibration time of $t_{\mathrm{eq}}=5\times10^3\tau'$, were discarded to allow the system to reach a steady state. The remaining $8\times10^7$ steps, corresponding
to $T_{\mathrm{data}}=4.0\times10^4\tau'$, were used for data collection and analysis. One integration step is counted after the update of all the ABP's and bead particles once.  We used 50 to 500 independent realizations for statistical averaging.

\section{Results \label{Results}}
\subsection{Shape deformations and dynamics of the ring}
\begin{figure*}[hbtp]
    \centering
    \includegraphics[width=1.0\linewidth]{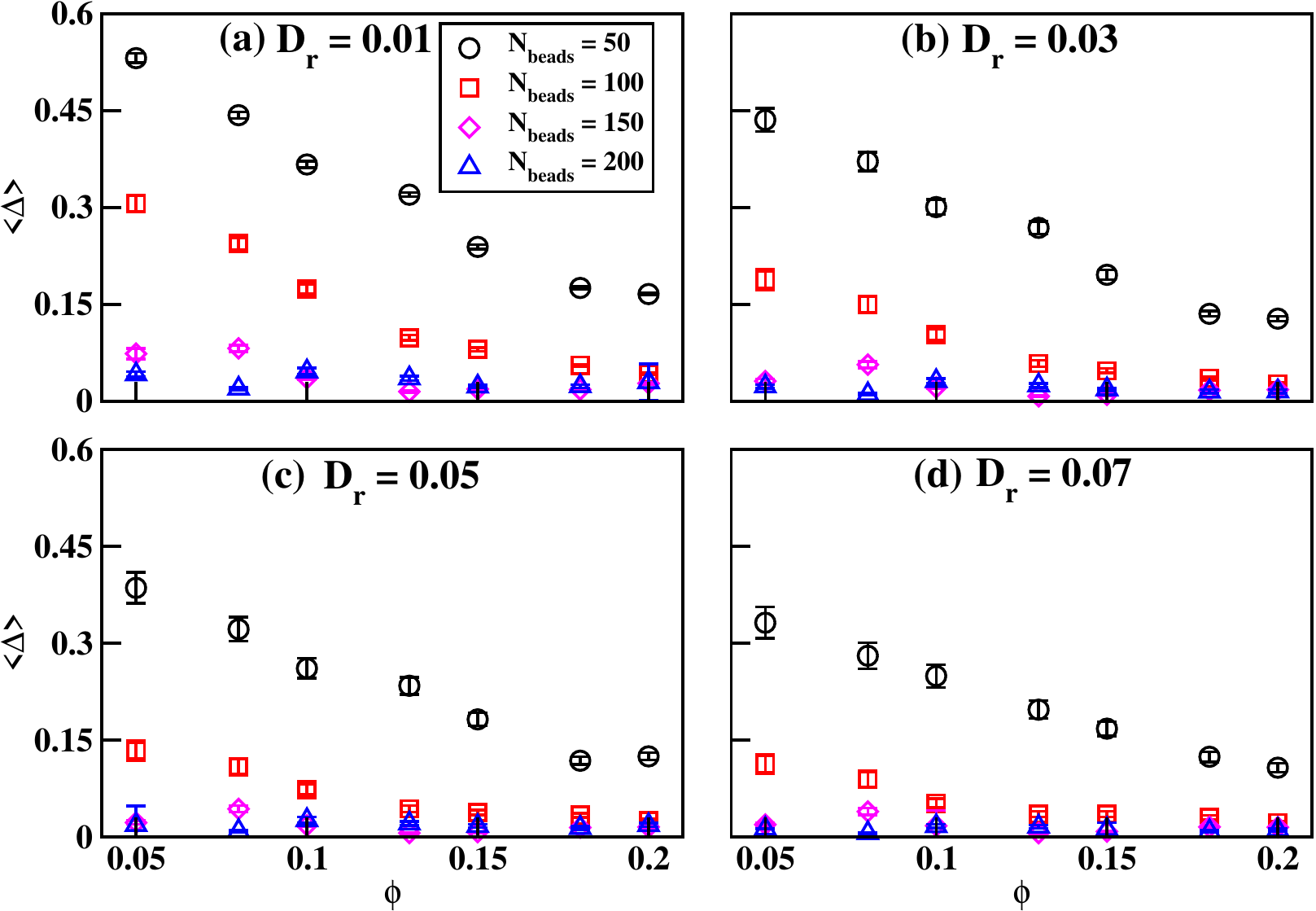}
     \caption{Variation of the mean asphericity $\langle \Delta \rangle$ {\it vs.} $\phi$ for different $D_r = 0.01$, $0.03$, $0.05$, and $0.07$. The different curves in each window in for different  $N_{\mathrm{beads}} = 50$, $100$, $150$, and $200$. Error bars represent the standard deviation obtained from independent realizations. }
    \label{fig:asph}
\end{figure*}

Fig.~\ref{fig:fig1}(i) illustrates ring configurations for different rotational diffusion coefficients, $D_r = 0.01, 0.03, 0.05$, and $0.07$, arranged from Fig.~\ref{fig:fig1}(i)(a-d) respectively. In these figures, the packing fraction is kept fixed at $\phi = 0.1$, and the ring consists of $N_{\rm{beads}} = 100$ beads. As $D_r$ increases, the orientations of the active particles randomize more rapidly, reducing the persistence of the propulsive forces exerted on the ring. Consequently, the active particles are unable to introduce significant force in a particular direction, resulting in a weaker ring deformation. As a result, ring shape fluctuations decrease, and the ring gradually approaches to a more circular configuration (\href{https://drive.google.com/drive/folders/1WoVYuwL4L9xFsH0lBXuVsrBFlv-HHHRj?usp=drive_link}{Movie-1}). This behavior is also evident from the particle trajectories shown in Fig.~\ref{fig:fig1}(i)(e-h). For small $D_r$, the active particles retain their orientations for longer, producing persistent motion that leads to larger ring deformations and extended trajectories. In contrast, at larger $D_r$, the particles reorient more frequently, causing their propulsion directions to become rapidly randomized. This frequent reorientation reduces the persistence of the forces acting on the ring, thereby suppressing ring deformations and resulting in significantly shorter trajectories.

Fig.~\ref{fig:fig1}(ii) shows the variation of the packing fraction of the ABPs while keeping the ring size fixed at $N_{\rm{beads}} = 100$ and the rotational diffusion coefficient fixed at $D_r = 0.01$. From Fig.~\ref{fig:fig1}(ii)(a-d), the packing fraction increases progressively. At low packing fractions, collisions between the active particles and the ring are relatively infrequent, resulting in an uneven distribution of forces along the ring. Consequently, the ring exhibits pronounced local shape deformations. As the packing fraction increases, particle bead collisions become more frequent and are distributed more uniformly along the ring boundary (\href{https://drive.google.com/drive/folders/1tQO7eA6M7Ixg9NBf__V-Sz4cZ4KtBBLn?usp=drive_link}{Movie~2}). This behavior is also evident from the corresponding trajectory plots shown in Fig.~\ref{fig:fig1}(ii)(e-h), where the ring follows smoother trajectories with fewer abrupt directional changes for larger packing fractions,  Fig.~\ref{fig:fig1}(ii)(g-h). The more homogeneous distribution of active forces suppresses localized ring deformations, causing the ring to adopt a smoother, more circular shape. Overall, increasing the packing fraction drives the system from a regime characterized by irregular, localized ring deformations to one dominated by uniform particle--ring interactions, resulting in enhanced shape stability.

Further, Fig.~\ref{fig:fig1}(iii) illustrates the effect of varying the number of ring beads, $N_{\mathrm{beads}} = 50, 100, 150$ and $200$, while keeping the packing fraction fixed at $\phi = 0.10$ and the rotational diffusion coefficient fixed at $D_r = 0.01$. The configurations, labeled Fig.~\ref{fig:fig1}(iii)(a-d), are arranged by increasing $N_{\mathrm{beads}}$. As the number of beads increases, the ring exhibits progressively smaller shape deformations, the ring boundary becomes more finely discretized, enabling the forces exerted by the confined active particles to be distributed more uniformly along the boundary(\href{https://drive.google.com/drive/folders/1AaZcbVi3HnDI_2LduQCsqX_VqH7c4AKr?usp=drive_link}{Movie~3}). This trend is also evident in the corresponding center-of-mass trajectory plots as shown in Fig.~\ref{fig:fig1}(iii)(e-h). For larger $N_{\rm beads}$, the ring maintains its circular shape with time, and dynamics get suppressed, as can be seen from the trajectory plot in Fig.~\ref{fig:fig1}(ii)(h), as compared to the other three cases (iii)(e-g). The homogeneous distribution of active forces suppresses the ring deformations, allowing the ring to maintain a more circular shape throughout its motion. Overall, increasing the number of beads promotes a more uniform transmission of active forces along the ring, thereby enhancing ring stability and reducing shape fluctuations.

To quantify the shape of the ring, we measure the asphericity $\Delta(t)$, defined with the help of gyration tensor $\mathbf{Q}$ \cite{aronovitz1986universal} \begin{equation}
\mathbf{Q}(t)
=
\frac{1}{N_{\mathrm{beads}}}
\sum_{n=1}^{N_{\mathrm{beads}}}
\left( \mathbf{R}_{n}(t) - \mathbf{R}_{\mathrm{cm}}(t) \right)
\otimes
\left( \mathbf{R}_{n}(t) - \mathbf{R}_{\mathrm{cm}}(t)\right).
\label{eq:gyration_tensor}
\end{equation}
where $\mathbf{R}_{\mathrm{cm}} = \frac{1}{N_{\mathrm{beads}}}\sum_{i=1}^{N_{\mathrm{beads}}}\mathbf{R}_i$ denotes the center-of-mass position of the ring, with $\mathbf{R}_i$ representing the position vector of the $i$-th bead. We construct the $2 \times 2$ gyration tensor $\mathbf{Q}$ and compute its eigenvalues. From the eigenvalues, $\lambda_1$ and $\lambda_2$ of $\mathbf{Q}$, we calculate the asphericity;
\begin{equation}
\Delta(t) = \frac{(\lambda_1(t) - \lambda_2(t))^2}{(\lambda_1(t) + \lambda_2(t))^2},
\end{equation}
 
$\Delta (t) \sim 0$ corresponds to a circular shape and $\Delta (t) \sim 1$ shows the elongated shape. 

We first calculate how the asphericity $\Delta(t)$ changes with time for different realizations. We found that for a fixed $\phi$ and $N_{\mathrm{beads}}$, for small $D_r$ values, $\Delta(t)$ shows strong variations for different realizations. This is due to the small $D_r$ values, which leads to a longer reorientation time for the active particles. Starting from the homogeneous random distribution of ABP's inside the ring, at very early times the particles distribute themselves at various locations at the boundaries of the ring; once they are there, they take longer to reorient, and hence the shape of the ring is stable for some time. Due to that, over the different realisations, we find $\Delta(t)$ shows variation from a value between $ 0.1$ and $ 0.5$, as can be seen in the time series of $\Delta(t)$ for $2$ distinct  realisations for $(D_r, \phi, N_{\rm beads}) = (0.01-0.07, 0.1, 100)$ Fig.\ref{time}(a-d) in the Appendix \ref{Time Series}.\\
The probability distribution $P(\Delta)$ of the mean of $\Delta(t)$, where the mean is calculated over time, shows a clear bimodal structure for the above parameters, Fig. \ref{fig:probe}(a). Now, as we increase $D_r$, the reorientation time decreases, and ABP's can redistribute easily along the boundary of the ring; hence, the time series show more fluctuations over time compared to the ensembles. The bimodality in $P(\Delta)$ weakens, as can be seen from the time series plots in Fig.\ref{time}(a-d) in the Appendix \ref{Time Series} and Fig. \ref{fig:probe}(a).

In Fig.\ref{fig:probe}(b), shows the plot of $P(\Delta)$ for different $\phi$ at fixed $D_r = 0.01$ and $N_{\rm beads} = 100$. For small packing fraction $\phi = 0.05$, the ring contains very few ABPs. Such a small number of ABP's with a large reorientation time creates kinks along the boundary of the ring, as can be seen in the snapshot in Fig.\ref{fig:fig1}(ii)(a). This leads to more intermittent fluctuations in the time series of $\Delta(t)$ as shown in Fig.\ref{time}(e) in Appendix \ref{Time Series}. Hence the $P(\Delta)$ is broadly distributed over the $\Delta \in (0, 0.7)$.  As we increase $\phi$, more and more ABPs are more homogeneously distributed across the boundary of the ring, resulting in smooth curvature across the boundary, a shift in $P(\Delta)$ towards smaller $\Delta$  value and weaker fluctuations with time as well as ensembles, as shown in Fig.\ref{time}(f-h) in Appendix \ref{Time Series}.

Finally, in Fig.\ref{fig:probe}(c) we show the effect of variation of the ring size for fixed values $\phi$ and $D_r$ by varying $N_{\mathrm{beads}}$. For small rings $N_{\mathrm{beads}} = 50$ and $100$, the distribution $P(\Delta)$ shows fluctuations from $\Delta \sim 0.5-0.6$ to $0.1$, due to the smaller size. But as we move to larger rings, the boundary of the ring acts like a continuous ring, and the force due to the ABPs is evenly distributed along the ring's boundary. This leads to a more circular shape of the ring and a narrower distribution of $P(\Delta)$. Time series of $\Delta(t)$ also show weaker fluctuations with time as well as ensembles, as can be seen in Fig.\ref{time}(i-l) in Appendix  \ref{Time Series}. 
 
\begin{figure*}[hbtp]
     \centering
     \includegraphics[width=1.0\linewidth]{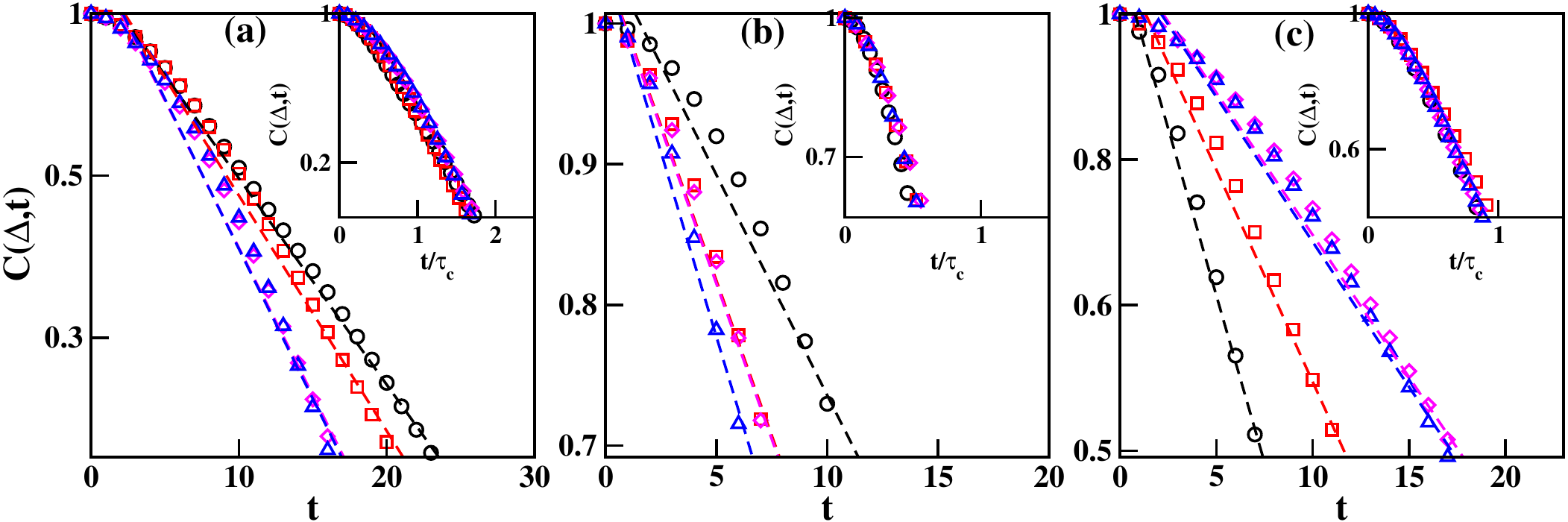}
     \caption {The time auto-correlation of the asphericity parameter $C(\Delta, t)$ {\it vs.} time $t$ on $\log$-y scale. 
     \textbf{(a)} At fixed $N_{\mathrm{beads}} = 100$, $\phi = 0.10$ and  for different  $D_r = 0.01$, $0.03$, $0.05$, and $0.07$. \textbf{(b)} At fixed $N_{\mathrm{beads}} = 100$, $D_r = 0.10$, and for different  $\phi = 0.05$, $0.10$, $0.15$, and $0.20$. \textbf{(c)} At  fixed $D_r = 0.01$, $\phi = 0.1$ and for different $N_{\mathrm{beads}} = 50$, $100$, $150$, and $200$. The insets in each window show the corresponding scaled correlation, where time is rescaled by the characteristic relaxation time $\tau_c$. All symbols have the same meaning as in Fig.~\ref{fig:probe}, and the dashed lines represent exponential fits.}
     \label{fig:correlation}
 \end{figure*}
Now we calculate the mean asphericity defined by $\langle \Delta(t)\rangle$, where $\langle...\rangle$ represents the mean over time and 50 independent realizations. In Fig.~\ref{fig:asph}(a-d), we show the variation of $\langle \Delta(t)\rangle$ as a function $\phi$ for different $D_r$. Different curves in each window are for different $N_{\rm beads}$.  For $N_{\rm beads} = 50$ and $100$, the mean asphericity decreases monotonically with increasing packing fraction, indicating that the ring progressively approaches a more circular shape as the density of confined active particles increases. Furthermore, for a fixed packing fraction, the mean asphericity also decreases with increasing $D_r$, reflecting the suppression of localized ring deformations due to the more rapid reorientation of the active particles. For large $N_{\mathrm{beads}} > 100$, the $\langle \Delta(t)\rangle$ is small and shows a very weak dependence on $\phi$.

Fig.~\ref{fig:correlation} shows the time auto-correlation of the shape fluctuation of the ring. The auto-correlation of the asphericity is defined as $C(\Delta, t) = \langle\delta \Delta(t_0) \delta \Delta(t+t_0)\rangle$, where $\langle...\rangle$, represents the mean over many independent realizations, reference time $t_0$.  $\delta \Delta(t) = \Delta(t) - \langle \Delta(t)\rangle$, is the fluctuation in asphericity.  In Fig.~\ref{fig:correlation}(a), the rotational diffusion coefficient $D_r$ is varied while keeping the packing fraction $\phi$ and the number of beads fixed at $N_{\rm{beads}} = 100$. It is observed that the autocorrelation function decays more rapidly, indicating more rapid shape fluctuations with increasing $D_r$. This behavior is attributed to the enhanced random reorientation of the active particles at higher rotational diffusion, which consequently reduces the relaxation time of the ring.

Similarly, for fixed $D_r = 0.1$ and $N_{\rm{beads}} = 100$, the $C(\Delta, t)$ decays faster with increasing packing fraction, see Fig.~\ref{fig:correlation}$(b)$, indicating rapid fluctuations of the ring as $\phi$ increases. Fig.~\ref{fig:correlation}(c) shows $C(\Delta, t)$ for different values of $N_{\rm{beads}}$ at fixed $D_r = 0.01$ and $\phi = 0.1$.  As $N_{\mathrm{beads}}$ increases, the autocorrelation curves shift upward and decay more slowly, indicating that larger rings have the size to move active particles from one side to the other side, and this helps to reduce the shape fluctuations. Consequently, the relaxation time increases with increasing $N_{\mathrm{beads}}$. For all of the above cases, the $C(\Delta, t)$ shows a nice scaling collapse when we rescale the time with the autocorrelation time ($\tau_c$) extracted from the early-time exponential fit of the correlation. In Fig.\ref{fig:correlation}(a-c) (insets) shows the plot of correlation vs. scaled time for different parameters.

In summary, confining active particles within a flexible ring produces behaviors fundamentally different from those seen in rigid-wall systems. At low particle densities, localized active forces drive strong, irregular deformations, while at higher densities the ring becomes smoother and more symmetric due to more uniformly distributed active particles at the ring's boundary. The ring’s flexibility introduces feedback: particle activity reshapes the boundary, and the evolving boundary in turn influences particle organization and motion.

\subsection{Dynamics of the ring}
\begin{figure*}[hbtp]
    \centering
    \includegraphics[width=1.0\linewidth]{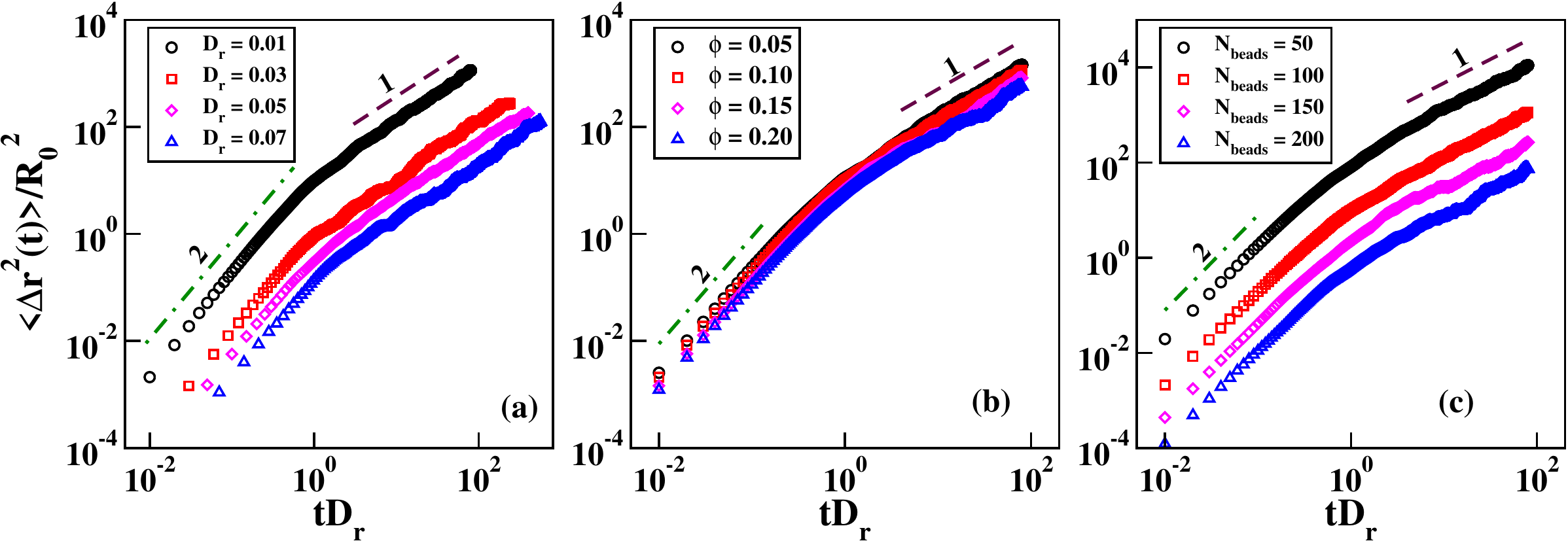}
       \caption{Mean-squared displacement (MSD) {\it vs.} time $t$ on  log--log scales. 
       \textbf{(a)} At fixed $N_{\mathrm{beads}} = 100$, $\phi = 0.10$ and  for different  $D_r = 0.01$, $0.03$, $0.05$, and $0.07$. \textbf{(b)} At fixed $N_{\mathrm{beads}} = 100$, $D_r = 0.10$, and for different  $\phi = 0.05$, $0.10$, $0.15$, and $0.20$. \textbf{(c)} At  fixed $D_r = 0.01$, $\phi = 0.1$ and for different $N_{\mathrm{beads}} = 50$, $100$, $150$, and $200$. The dotted-dashed and dashed lines show the lines with slope 2 and 1, respectively.}
    \label{fig:msd}
\end{figure*}
Till now, we have focused on the structural characteristics of the ring. In this subsection, we will examine how the ring dynamics respond to shape fluctuations. The activity of the confined self-propelled particles also induces translational motion of the ring. To characterize the translational motion, we calculate the mean-squared displacement (MSD) of the center of mass of the ring as defined by the $\Delta r^2(t) = \langle [{\bf R}_{\rm cm}(t) - {\bf R}_{\rm cm}(0)]^2\rangle$, where $\langle...\rangle$ is mean over many realizations and reference times $t_0$, which is marked as $0$ in the definition of $\Delta r^2(t)$. In Fig.~\ref{fig:msd}(a-c) we show the $\Delta r^2(t)$ vs. time $t$  for different $D_r$, $\phi$ and  $N_{\mathrm{beads}}$ respectively.

For all the parameters, at short times, the MSD curves display a ballistic regime characterized by $\langle \Delta r^2(t) \rangle \sim t^2$, reflecting the persistent motion generated by the active particles enclosed within the ring. As time increases, the system undergoes a crossover from ballistic to diffusion.

At late times, all MSD curves eventually approach a diffusive regime characterized by $\langle \Delta r^2(t) \rangle \sim t$. This transition reflects the gradual loss of directional persistence due to the continuous reorientation and redistribution of active particles inside the ring. As a result, the center-of-mass motion becomes effectively random, although the effective diffusion constant very much depends on $D_r$, $\phi$, and $N_{\mathrm{beads}}$. The emergence of long-time diffusive behavior demonstrates that the active ring behaves as an effective Brownian particle at sufficiently long time scales \cite{wang2019shape,le2022encapsulated,takatori2020active,vutukuri2020active,iyer2022non,paoluzzi2016shape}.
\begin{figure*}[hbtp]
    \centering
    \includegraphics[width=1.0\linewidth]{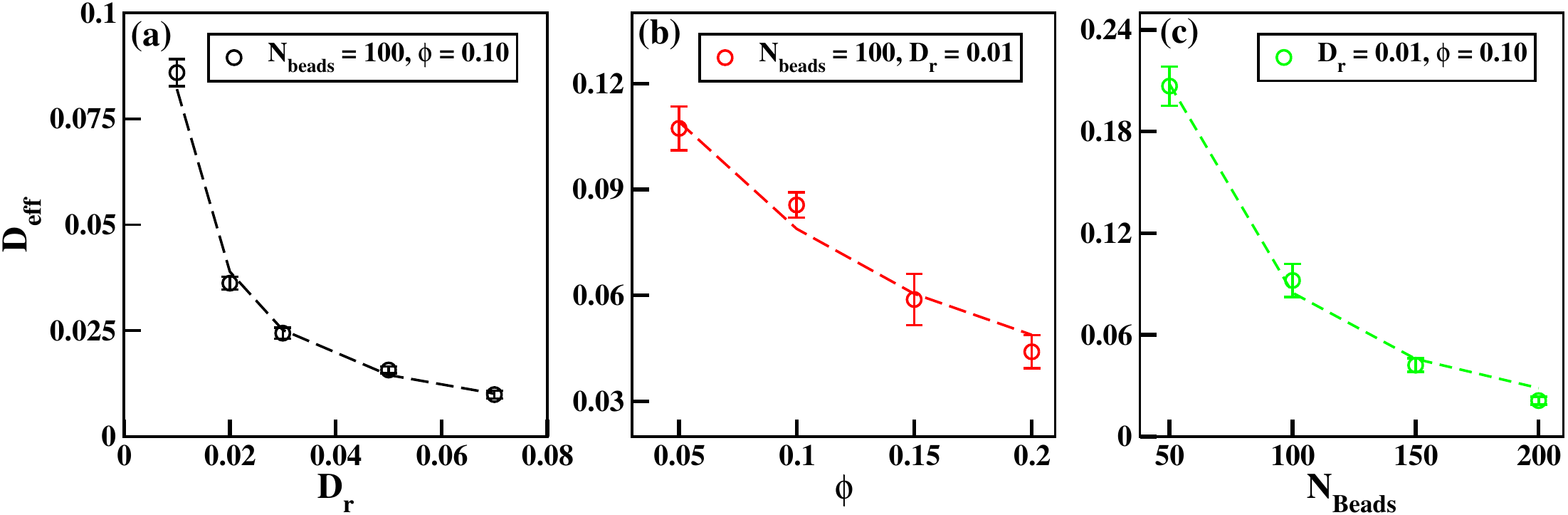}
    \caption{Effective diffusion coefficient, $D_{\mathrm{eff}}$  of the ring as a function of key system parameters. \textbf{(a)} $D_{\mathrm{eff}}$ {\it vs.} $D_r$ for fixed $N_{\mathrm{beads}}=100$ and $\phi=0.10$. The dashed line shows the fitting of the data point with $1/D_r$. \textbf{(b)} $D_{\mathrm{eff}}$ {\it vs.} $\phi$ for fixed $N_{\mathrm{beads}}=100$ and $D_r=0.01$.  \textbf{(c)} $D_{\mathrm{eff}}$ {\it vs.} $N_{\mathrm{beads}}$, for fixed $D_r=0.01$ and $\phi=0.10$.  In (b) and (c), dashed lines are the data points fitted with the expression of $D_{\mathrm{eff}}$ as given in Eq.\ref{eq:diff}.}
    \label{fig:diffusion}
\end{figure*}

Next we calculate the dependence of effective diffusivity $D_{\mathrm{eff}} = \lim_{t \to \infty} \frac{\Delta r^2(t)}{4 t}$, where $\lim_{t \rightarrow \infty}\frac{\Delta r^2(t)}{4 t}$, means the mean value of $\frac{\Delta r^2(t)}{4 t}$ at late times. 
This effective diffusivity $D_{\mathrm{eff}}$ provides a quantitative measure of the efficiency of active transport of the ring as a single active particle. As shown  in Fig.~\ref{fig:diffusion}(a), $D_{\mathrm{eff}}$  decreases  with increasing $D_r$. Since the persistence length scales as $l=v_0/D_r$, smaller values of $D_r$ allow active particles to maintain their orientations for longer times, thereby enhancing the persistence of the net propulsion force. This leads to larger ballistic displacements and higher long-time diffusivities. In contrast, increasing $D_r$ causes more rapid reorientation of the active particles, shortening the persistence time and suppressing directed motion. As a result, the effective diffusivity decreases as $1/D_r$,  with increasing rotational diffusion. In Fig. \ref{fig:diffusion}(a), the data points are  $D_{\mathrm{eff}}$ calculated from the numerical simulation, and the dashed lines are the data points fitted with the expression of $D_{\mathrm{eff}}$ as given in Eq.\ref{eq:diff}.

We further plot the $D_{\mathrm{eff}}$ with increasing packing fraction $\phi$ as shown in Fig. \ref{fig:diffusion}(b). Again, on increasing $\phi$, the $D_{\mathrm{eff}}$ decreases monotonically.  A similar trend is observed with increasing ring size $N_{\mathrm{beads}}$. Larger rings possess greater area and experience larger effective drag, requiring stronger collective forces to achieve the same level of translational motion. Consequently, the center-of-mass displacement becomes less efficient, leading to a reduction in the long-time diffusion coefficient. This size dependence highlights the important role of ring geometry in regulating active transport.

We also fit the MSD obtained from our simulations in Fig. \ref{fig:msd}(a-c) using the theoretical expression given in Eq.~\ref{msd_cm} (see \ref{Mathematical Description of MSD}). The effective diffusivity is calculated as $D_{\mathrm{eff}}$ in Eq.\ref{eq:diff}. \\
\begin{figure*}[hbtp]
    \centering
    \includegraphics[width=1.0\linewidth]{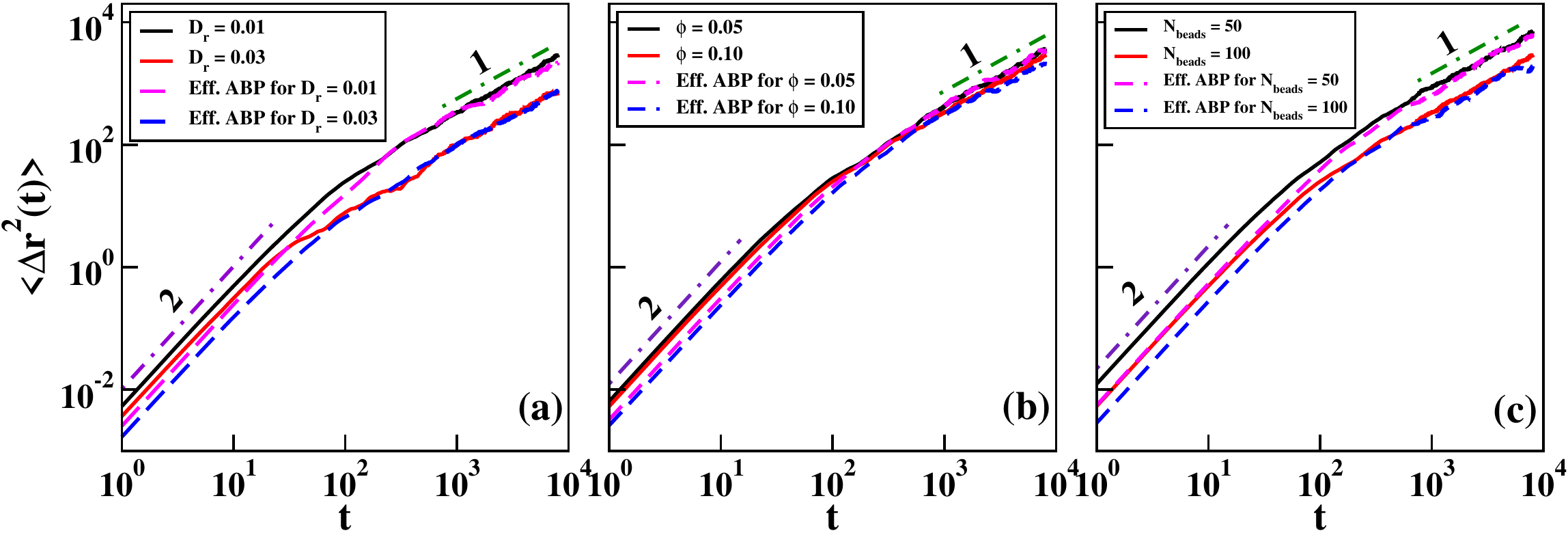}
    \caption{MSD of a ring as an effective ABP in comparison with a full ring, with respect to time $t$, plotted on log--log scales. \textbf{(a)} For different $D_r = 0.01, 0.03$ at fixed packing fraction $\phi = 0.10$ and $N_{\mathrm{beads}} = 100$.\textbf{(b)} For different  $\phi = 0.05, 0.10$ at fixed $D_r = 0.01$ and $N_{\mathrm{beads}} = 100$. \textbf{(c)} For different  $N_{\mathrm{beads}} = 50, 100$ at fixed $D_r = 0.01$ and $\phi = 0.10$. }
    \label{fig:msd_eff}
\end{figure*}
Overall, the dynamics of active rings are governed by a complex interplay between particle activity, ring deformability, confinement, and crowding. While ring fluctuations continuously modify the instantaneous shape and propulsion characteristics of the ring, the system consistently exhibits a crossover from ballistic to diffusive motion. The observed dependence of the diffusivity on $\phi, D_r$, and $N_{\mathrm{beads}}$ demonstrates how both internal particle dynamics and ring mechanics regulate collective active forces. These results establish active rings as emergent self-propelled objects whose transport properties can be tuned through activity, confinement, and ring size.

\section{ring as an effective Active Brownian Particle (ABP) \label{ring as an effective Active Brownian Particle(ABP)}}
To further characterize the ring dynamics, we map the motion of the ring with an effective Active Brownian Particle (ABP) framework. For this purpose, we simulate a single effective ABP using the parameters obtained from the ring dynamics. The particle move with an effective self-propulsion speed, $v_{\mathrm{eff}} = \sqrt{2 D_{\mathrm{eff}}/t_c}$, where $t_c$ is obtained by calculating the crossover time from a ballistic to diffusive regime in MSD. The position of the  particle evolves according to

\begin{equation} \frac{d\mathbf{r}}{dt} = {v_{\mathrm{eff}}}\mathbf{p}(t) + \sqrt{2D_{\mathrm{eff}}}\boldsymbol{\zeta}_r. 
\end{equation}
The orientation of the particle is given by  
\begin{equation}
\frac{d\boldsymbol{\theta}}{dt}
=
\sqrt{2\nu_{r}}\boldsymbol{\zeta}_\theta.
\end{equation}
where the direction vector is $\mathbf{p}(t) = \bigl(\cos \theta(t), \, \sin \theta(t)\bigr)$. Here $\nu_{r} = t_c^{-1}$, and $D_{\mathrm{eff}}$ represents the effective rotational and translational diffusion coefficients for the ring as an effective self-propelled particle. The term \(\zeta_r(t)\) , \(\zeta_\theta(t)\) is an independent Gaussian white-noise stochastic processes satisfying $\langle \zeta_r(t) \rangle = 0$, $\langle \zeta_\theta(t) \rangle = 0$ and, $\langle \zeta_r(t)\zeta_r(t') \rangle = \delta(t-t')$, $\langle \zeta_\theta(t)\zeta_\theta(t') \rangle = \delta(t-t')$. Here, \(\delta(t-t')\) denotes the Dirac delta function.

We calculate the $v_{\mathrm{eff}}$, $\nu_{r}$, and $D_{\mathrm{eff}}$ for different parameters from the simulation of the full ring. Further, these effective values are fed to calculate the mean square displacement of the ring as an effective self-propelled particle.
As shown in Fig.~\ref{fig:msd_eff}, we compare the MSD of the full ring and the ring as an effective self-propelled particle for different parameters: ($D_r = 0.01, 0.03$), ($\phi = 0.05, 0.10$), and ($N_{\mathrm{Beads}}= 50, 100$). 
We see a good agreement between the two cases, which demonstrates that the ring dynamics are well captured by an effective Active Brownian Particle description. This result indicates that, despite the underlying complexity of the active particles-beads interactions, the effective dynamics can be understood in terms of persistent active propulsion combined with stochastic rotational diffusion.

\section{Conclusion \label{Conclusion}}
We systematically explore the dynamics and shape deformation of the ring by varying three key control parameters: the rotational diffusion coefficient, the packing fraction of active particles, and the number of beads forming the ring. We find that all three parameters strongly influence the conformational dynamics and shape fluctuations of the deformable ring. In particular, the ring exhibits pronounced temporal fluctuations in its shape, continuously evolving between elongated and nearly circular configurations. At low packing fractions and small rotational diffusion coefficients, the active particles tend to remain localized for longer times, resulting in a more heterogeneous spatial distribution along the ring boundary and consequently larger and more persistent shape deformations. As the packing fraction is increased, the larger number of active particles facilitates a more homogeneous redistribution of particles within the confined region and along the boundary, thereby suppressing localized deformations and driving the ring toward a more circular configuration. Similarly, increasing the rotational diffusion coefficient enhances the reorientation of the active particles, allowing them to redistribute more rapidly and reducing the persistence of localized deformations. The size of the deformable ring also plays an important role: increasing the number of beads provides a finer representation of the boundary and promotes a more uniform spatial distribution of the active particles along the ring, resulting in reduced asphericity and weaker shape fluctuations. Thus, although the three parameters affect the system through different aspects of particle dynamics and ring geometry, they collectively control the degree of deformation by regulating how uniformly the active particles are distributed within and along the deformable boundary. This demonstrates that the conformational dynamics of flexible confinement can be systematically controlled by tuning both the dynamical properties of the active particles and the structural characteristics of the ring.\\

Beyond these shape fluctuations, the ring itself exhibits emergent dynamics and behaves as an effective active object. The collective motion of the confined self-propelled particles gives rise to persistent random motion of the ring, with its translational dynamics depending strongly on the properties and spatial organization of the enclosed particles. Thus, the shape fluctuations and internal particle dynamics are reflected in the motion of the ring as a whole, demonstrating how the activity of the confined particles is transferred to the flexible boundary and gives rise to emergent dynamics at the scale of the entire ring.
These results demonstrate that internal activity alone, without any external guidance, can generate collective and directed motion in a deformable confined system.\\

Interestingly, this behavior bears similarities to cell migration in biological systems, where cells develop polarity and perform a persistent motion. Further, it can be utilized to develop simple models useful for wound healing \cite{li2013collective} or dynamics in response to chemical cues \cite{petrie2009random,liebchen2018synthetic}.  \\

The current study has focused on the shape deformation and dynamics of a single ring and found that by tuning the properties and density of confined particles, the center of mass of the ring performs persistent to diffusive motion. That itself implies an emergent asymmetry in the ring geometry.
It would be interesting to investigate whether this asymmetry can induce effective orientational ordering among multiple rings and, ultimately, lead to their collective flocking. Through this work, we provide a simple physical framework for understanding how microscopic activity and confinement can combine to produce shape adaptation, polarization, and migration reminiscent of biological systems.\\

\section{ACKNOWLEDGMENTS \label{ACKNOWLEDGMENTS}} A K and S M acknowledge the computational support provided by PARAM Shivay under the National Supercomputing Mission, Government of India, hosted at Indian Institute of Technology (BHU) Varanasi. S M further acknowledges financial support from the Department of Science and Technology – Science and Engineering Research Board under the grants CRG/2021/006945,  MTR/2021/000438 and ANRF/ARG/2025/008220/PS.

\section{DATA AVAILABILITY STATEMENT \label{DATA AVAILABILITY STATEMENT}}
All data that support the findings of this study are included within the article.

\onecolumngrid
\appendix
\section{Time Series of Asphericity\label{Time Series}}
\begin{figure*}[hbtp]
    \centering
    \includegraphics[width=1.0\linewidth]{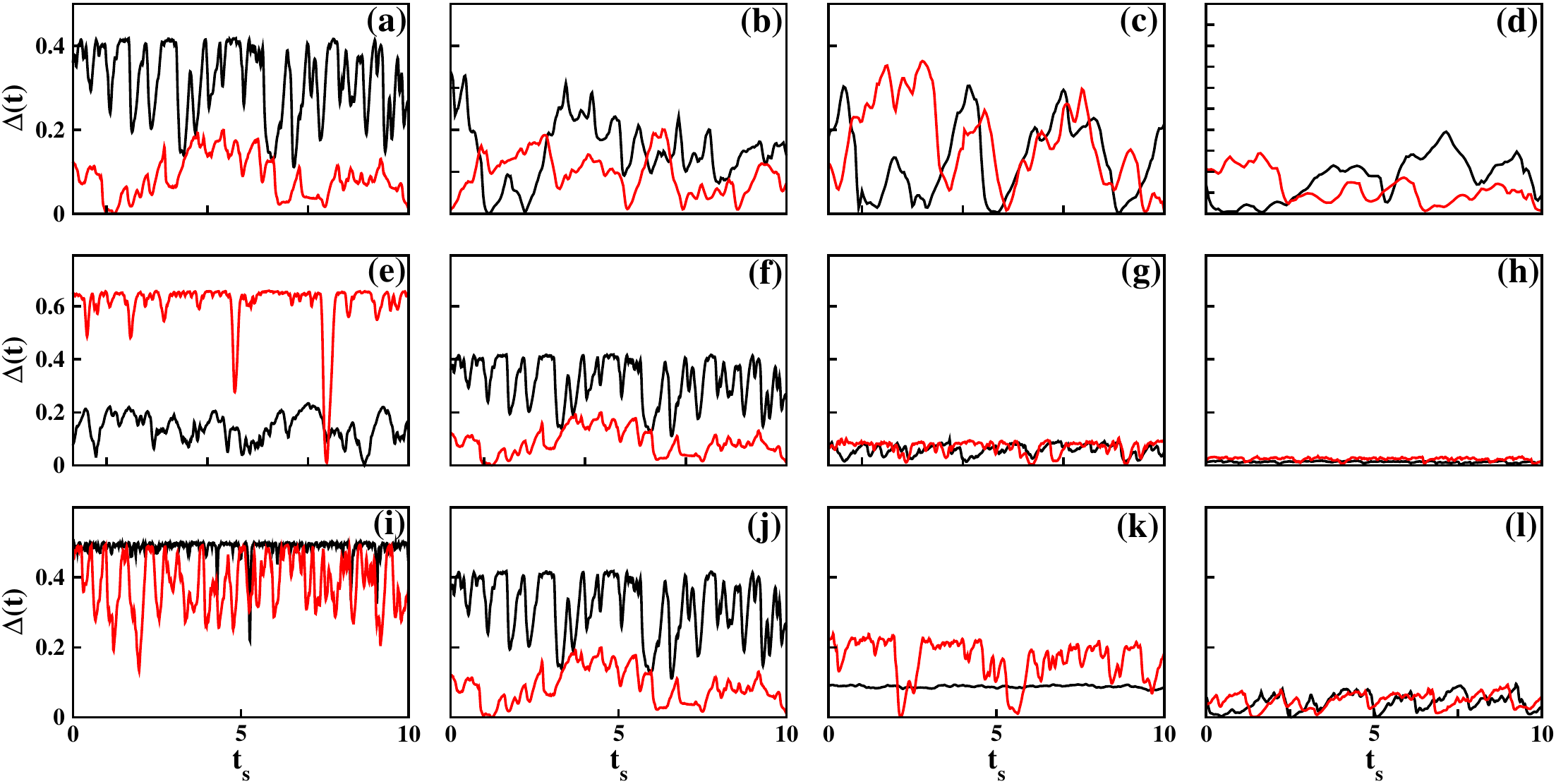}
    \caption{Time evolution of the asphericity parameter $\Delta(t)$ over the steady-state in a fixed scaled time interval $t_s = t D_r = 10$. (a-d) is for  fixed $\phi=0.10$ and  $N_{\mathrm{Beads}}=100$, and  varying $D_r = 0.01, 0.03, 0.05$ and $0.07$ respectively.  (e-h) is for different $\phi = 0.05, 0.1, 0.15$ and $0.2$ respectively at fixed  $D_r=0.01$ and $N_{\mathrm{Beads}}=100$. (i-l) is by varying $N_{\mathrm{beads}} = 50, 100, 150$ and $200$ respectively at fixed $\phi = 0.1$ and $D_r = 0.01$.  In each window, for all set of parameters, the two plots are for two distinct realizations.}
    \label{time}
\end{figure*}

Fig.~\ref{time}(a-l) shows the time evolution of $\Delta(t)$ for a time interval $t D_r = 10$ (in the steady state) for different parameters for two distinct realizations. Fig.~\ref{time}(a-d) is  for different values of $D_r = 0.01, 0.03, 0.05$ and $0.07$ respectively at fixed $\phi = 0.1$, $N_{\mathrm{beads}} = 100$. For small $D_r = 0.01$, Fig.~\ref{time}(a), the ring exhibits fluctuations in its shape while maintaining a relatively well-defined range of asphericity. As $D_r$ increases, Fig~\ref{time}(b-c), the average asphericity decreases. This behavior can be attributed to the rapid reorientation of the active particles, and consequently suppresses large shape deformations.

The Fig.~\ref{time}(e-h) shows  $\Delta(t)$, for different values of the packing fractions, $\phi = 0.05, 0.1, 0.15$ and $0.2$ respectively and at fixed $D_r 0.01$ and $N_{\mathrm{beads}} = 100$. For small $\phi = 0.05$, as shown in Fig.~\ref{time}(e), the ring exhibits temporal fluctuations in its shape while maintaining a relatively well-defined range of asphericity. As the packing fraction increases, the active particles become more uniformly distributed within the ring, leading to a more homogeneous distribution of forces and thereby reducing localized deformations of the ring. Consequently, the mean asphericity decreases, indicating that the ring becomes progressively more symmetric with increasing $\phi$, as illustrated by the representative snapshots in Fig.~\ref{time}(f-h).

Further $\Delta(t)$ for varying $N_{\mathrm{beads}} = 50, 100, 150$ and $200$ is shown in Fig.~\ref{time}(i-l) respectively at fixed $\phi = 0.1$ and $D_r = 0.01$. For small $N_{\mathrm{beads}} = 50$, as shown in Fig.~\ref{time}(i), the ring exhibits temporal fluctuations in its shape while maintaining a relatively well-defined range of asphericity. As $N_{\mathrm{beads}}$ increases, the ring size becomes larger, and, at fixed packing fraction $\phi$, the active particles have more space to redistribute within the ring. This increased spatial freedom promotes a more uniform distribution of active particles. As a result, localized deformations are suppressed, and the mean asphericity decreases, indicating that the ring becomes progressively more symmetric with increasing $N_{\mathrm{Beads}}$. The corresponding changes in ring morphology are illustrated by the representative plots in Fig.~\ref{time}(j-l).

\section{Mean square displacement (MSD) of ring as an effective SPP \label{Mathematical Description of MSD}}
Here we follow a similar description for the dynamics of the ring as given in \cite{paoluzzi2016shape}. We can write an effective dynamics of the center of mass of the ring by considering that it is made up of $N_{\rm{beads}}$ and $N_{\rm{active}}$ active particles. For simplicity, we assume that the mobility of the ring is the same as the mobility of individual particles. Since the bead particles form the boundary of the ring, we can also say that, due to the balance of forces, the effective force responsible for the ring dynamics is coming only from the active particles enclosed within it. Hence equation of motion of the beads center of mass ${\bf R}_{cm}(t)$ can be written as 
\begin{equation}
\dot{\mathbf{R}}_{\rm cm}
=
\frac{\mu}{N_{\mathrm{beads}} + N_{\mathrm{active}}}
\sum_i \mathbf{F}_{\mathrm{active}}.
\end{equation}
 $\mathbf{F}_{\mathrm{active}}$ represents the total propelling force exerted by the active particles.   Consequently, the center of mass moves with an effective mobility $\frac{\mu}{N_{\mathrm{beads}} + N_{\mathrm{active}}}$ under the action of the total propelling force. The corresponding velocity--velocity correlation function is given by

\begin{equation}
\left\langle
\dot{\mathbf{R}}_{\mathrm{cm}}(0)
\cdot
\dot{\mathbf{R}}_{\mathrm{cm}}(t)
\right\rangle
=
\frac{\mu^2}{(N_{\mathrm{active}}+N_{\mathrm{beads}})^2}
\sum_{\substack{i,j=1 \\i\neq j}}^{N_{\mathrm{active}}}
\left\langle
\mathbf{F}_i(0)
\cdot
\mathbf{F}_j(t)
\right\rangle .
\label{eq:velocity_correlation}
\end{equation}
For active particles, propelling forces only reorient due to rotational diffusion and are therefore uncorrelated, so that 
\begin{equation}
\left\langle \hat{\mathbf{n}}(0) \cdot \hat{\mathbf{n}}(t) \right\rangle
= e^{-D_r t}
\end{equation}
\begin{equation}
   \left\langle
\mathbf{F}_i(0)\cdot\mathbf{F}_j(t)
\right\rangle
= {F}^2 e^{-D_r t}{\delta_{ij}} = \frac{v_0^2}{\mu^2} e^{-D_r t}{\delta_{ij}}
\end{equation}
\begin{equation}
\sum_{i,j=1}^{N_{\mathrm{active}}}
\left\langle
\mathbf{F}_i(0)\cdot\mathbf{F}_j(t)
\right\rangle
=
N_{\mathrm{active}}
\left\langle
\mathbf{F}(0)\cdot\mathbf{F}(t)
\right\rangle
=
{N_{\rm{active}}}\frac{v_0^2}{\mu^2} e^{-D_r t},
\label{eq:force_correlation}
\end{equation}

so Eq.~\ref{eq:velocity_correlation} becomes 
\begin{equation}
\left\langle
\dot{\mathbf{R}}_{\mathrm{cm}}(0)
\cdot
\dot{\mathbf{R}}_{\mathrm{cm}}(t)
\right\rangle
=
\frac{\mu^2}{(N_{\mathrm{active}}+N_{\mathrm{beads}})^2} \frac{v_0^2}{\mu^2} e^{-D_r t}
\label{average}
\end{equation}

The mean square displacement (MSD) is obtained by a double time
integration of Eq.~\eqref{average}, yielding
\begin{equation}
\left\langle
\Delta \mathbf{R}_{\mathrm{cm}}^2(t)
\right\rangle
=
\frac{v_0^2 N_{active}}{(N_{\mathrm{active}}+N_{\mathrm{beads}})^2}
\int_0^t dt'
\int_0^{t'} dt''
\,e^{-D_r(t'-t'')}
\end{equation}
so after integrating over time 
\begin{equation}
\left\langle
\Delta \mathbf{R}_{\mathrm{cm}}^2(t)
\right\rangle
=
4\frac{v_0^2 N_{\mathrm{active}}}
{2D_r^{{2}}\left(N_{\mathrm{beads}}+N_{\mathrm{active}}\right)^2}
\left(
D_r t - 1 + e^{-D_r t}
\right)
\label{msd_cm}
\end{equation}
which gives 
\begin{equation}
    D_{\mathrm{eff}} = \frac{v_0^2 N_{\mathrm{active}}}{2D_r(N_{\mathrm{beads}}+N_{\mathrm{active}})^2}
    \label{eq:diff}
    \end{equation}  
    Further, we can write $N_{\mathrm{active}}$ in terms of packing fraction as defined in the model section and also substitute the value of $N_{\mathrm{beads}}$.

\section{Detail of Movies \label{Movies}}
All movie files are named according to the format ``{\it value of $D_{r}-\phi-N_{\mathrm{beads}}$}.mp4''.\\
\textbf{Movie 1}- Movie 1 shows the ring motion at different Rotational Diffusion Coefficients.\\
Link-\textcolor{blue}{\url{https://drive.google.com/drive/folders/1WoVYuwL4L9xFsH0lBXuVsrBFlv-HHHRj?usp=drive_link}}

\textbf{Movie 2-} Movies 2nd shows the ring motion at different packing fractions. \\  
Link - \textcolor{blue}{\url{https://drive.google.com/drive/folders/1tQO7eA6M7Ixg9NBf__V-Sz4cZ4KtBBLn?usp=drive_link}}

\textbf{Movie 3-} Movie 3rd shows the ring motion at different No. of bead particles.\\
Link - \textcolor{blue}{\url{https://drive.google.com/drive/folders/1AaZcbVi3HnDI_2LduQCsqX_VqH7c4AKr?usp=drive_link}}

\twocolumngrid
\bibliography{citation}
\end{document}